\documentclass[11pt,a4papers,floats]{article}
\usepackage{amsfonts,color,epsfig}
\usepackage{multirow}
\usepackage[multiple]{footmisc}
\usepackage{booktabs}
\renewcommand{\theequation}{A-\arabic{equation}}
\newcommand{\nc}{\newcommand}

\newcommand{\rnc}{\renewcommand}
\renewcommand{\thefootnote}{\fnsymbol{footnote}}
\makeatletter
\rnc{\theequation}{\thesection.\arabic{equation}}
\@addtoreset{equation}{section}
\makeatother
\nc{\fig}[5] {
\begin{figure}[!htbp]
    \begin{center}
    \leavevmode
    \centerline{
        \includegraphics[width=#1, height=#2]{#3}
        }
    \caption[]{#4}
    \label{#5}
    \end{center}
\end{figure}}
\nc{\figs}[8]{
\begin{figure}[!htbp]
    \begin{center}
    \leavevmode
    \centerline{
        \includegraphics[width=#1, height=#2]{#3}
        \includegraphics[width=#4, height=#5]{#6}
        }
    \caption[]{#7}
    \label{#8}
    \end{center}
\end{figure}}

\begin{document}
\begin{flushright}
{\small \tt arXiv:0811.2632\\
\tt [hep-th]}
\end{flushright}
\vspace{5mm}
\begin{center}
{\Large {\bf Absorption Cross Section in Warped AdS$_3$ Black Hole}}\\[10mm]

{John J. Oh$^{a}$\footnote{Email: johnoh@nims.re.kr} and Wontae Kim$^{b,c}$\footnote{Email: wtkim@mail.sogang.ac.kr}}\\[10mm]

{\small {${}^{a}$\it Division of Interdisciplinary Mathematics, 
National Institute for Mathematical Sciences,\\ Daejeon 305-340, South Korea\\[0pt]
${}^{b}$ \it {Department of Physics, Sogang University, C.P.O. Box 1142, Seoul 100-611, South Korea} \\[0pt]
${}^{c}$ \it Center for Quantum Spacetime, Sogang University, Seoul 121-742, South Korea      
}}
\end{center}
\vspace{4mm}
\begin{abstract}
The absorption cross section is studied in the low-frequency region 
for a propagating scalar field under the warped AdS$_3$ black hole
background in the  { cosmological} topologically massive gravity. 
It can be shown that the absorption cross section is  { unexpectedly} deformed 
by the gravitational Chern-Simons term, which is proportional to 
the scattering area of black hole with an additional contribution 
depending on the combinations of left-moving and right-moving temperatures.
It means that the cross section is larger than the area in spite of
the s-wave limit.
Finally, we discuss the left-right quasinormal modes for the scalar
perturbation in this black hole. 
\end{abstract}
\vspace{5mm}


\vspace{2cm}

\hspace{11.5cm}{Typeset Using \LaTeX}
\newpage
\renewcommand{\thefootnote}{\arabic{footnote}}
\setcounter{footnote}{0}
\section{Introduction}\label{sec:intro}
There has been much attention to three-dimensional gravity for many years
because 
it certainly offers potential insights into quantum gravity. 
In either Einstein-Hilbert action with cosmological constant or 
gravitational Chern-Simons theory (GCS) in three-dimensions, 
there are no propagating degrees of freedom in the bulk even though 
there are asymptotically AdS$_3$ black hole solution
\cite{Banados:1992wn}. However, the combined theory called as ``{cosmological} topologically massive
gravity'' (TMG) {has} a single massive graviton mode in the bulk \cite{Deser:1981wh}. 
On the other hand, it is easy to check that the solution to this theory is the BTZ black
hole as a trivial class of solutions \cite{Banados:1992wn,Banados:1992gq}.

Recently, the three-dimensional Einstein gravity 
with a negative cosmological constant and the GCS term with
coefficient $1/\mu$ 
has been considered for AdS$_3$ black hole
\cite{Li:2008dq}, {whose action is
\begin{equation}
\label{eq:tmg}
S_{TMG} = \frac{1}{2\kappa^2} \int d^3 x \sqrt{-g}
\left(R+\frac{2}{\ell^2}\right)+ \frac{1}{4\kappa^2\mu} \int d^3 x
\sqrt{-g} \tilde{\epsilon}^{\lambda\mu\nu} \Gamma_{\lambda\sigma}^{\rho}\left[\partial_{\mu} \Gamma_{\nu\rho}^{\sigma} + \frac{2}{3}\Gamma_{\mu\tau}^{\sigma}\Gamma_{\nu\rho}^{\tau}\right],
\end{equation}
where $G_{N}$ is the three-dimensional Newton's constant with
$\kappa^2 = 8\pi G_{N}$, $\tilde{\epsilon}^{\lambda\mu\nu}$ is a
tensor defined 
by ${\epsilon}^{\lambda\mu\nu}/\sqrt{-g}$ with ${\epsilon}^{012}=1$, 
and $\mu\equiv 3\nu/\ell$ is the dimensionless coupling constant. If we choose a positive sign for the Einstein-Hilbert term as above, the black holes have positive energy for $\mu\ell >1$ while massive gravitons have negative energy. However, if we choose a ``wrong'' sign, the massive graviton has positive energy\footnote{See Refs. \cite{Carlip:2008jk, Li:2008yz} for the negative Newton's constant.}
}

  {It has been shown that} there are unstable and inconsistent vacua for generic $\mu$ 
due to the massive graviton with negative energy in the bulk \cite{Li:2008dq}. 
However, at the chiral point of $\mu\ell =1$, 
it is found that the stable vacua suggests the existence of the
consistent chiral gravity and
its boundary CFT has a purely
right-handed 
chirality with a central charge of $c_{R} = 3\ell /G$. 
Moreover, it has been argued that the energy for gravitons vanishes at this critical point \cite{Li:2008dq} and thus that the massive graviton disappears. 
  {However, it was argued that the logarithmic graviton mode with negative energies, arising from the degeneracy of the left and massive branches, persists at the chiral point \cite{Grumiller:2008qz}, in which it is also shown that the boundary stress tensor is chiral while the dual CFT is logarithmic (not chiral).
In particular, it was argued that the restriction to Brown-Henneaux boundary condition does not remove the descendants of the logarithmic mode \cite{Giribet:2008bw} and the mode is consistent with the asymptotically AdS spacetimes \cite{Grumiller:2008es}.}
This issue has been disputed in several papers,
leading to an intensive discussion \cite{Grumiller:2008qz,Giribet:2008bw,Grumiller:2008es,Carlip:2008jk,Li:2008yz,Carlip:2008eq,Sachs:2008gt,Grumiller:2008pr,Strominger:2008dp,Carlip:2008qh,Park:2008yy}. 

On the other hand, 
aside from all unstable vacua of AdS$_3$ black holes, 
``warped AdS$_3$ vacua'' have been found 
for every $\mu$ \cite{Anninos:2008fx}.
The warped AdS$_3$ geometry can be viewed 
as a fibration of the real line with a constant warp factor over
AdS$_2$, 
which reduces the $SL(2,R)_{L}\times SL(2,L)_{R}$ isometry group 
to $SL(2,R)\times U(1)$.
One of the solutions, which is free from
naked closed timelike curves (CTCs), is
the spacelike stretched black hole. 
The spacelike stretched black hole solution has been  
also studied in Ref. \cite{Bouchareb:2007yx, Moussa:2008sj}.
For this geometry, it was conjectured that the TMG with 
$\nu= \frac{\mu\ell}{3}>1$ is dual to the (1+1)-dimensional 
boundary CFT with an asymmetric central charges 
of $c_{R} = (5\nu^2+1)\ell/G\nu(\nu^2+3)$ and $c_{L} =
4\nu\ell/G(\nu^2+3)$, which is still being disputed
\footnote {{See Ref. \cite{Compere:2008cv} for the alternative viewpoint on this issue, in which the central charge with the opposite sign is derived, yielding the instability of the warped geometry.}}.

In this paper, we would like to study the behavior of a scalar field
under the spacelike stretched black
hole background and compute its absorption cross section 
using coefficients matching. 
The essential GCS effect in the metric 
is that the rotational region is the
whole spacetime since the stationary observer is possible only at the
infinity. {This rotational effect in the metric} can modify the conventional absorption cross section 
when we consider the scattering modes.
The outline of our paper is as follows.
In Sec. \ref{sec:tmg}, we briefly review the  {cosmological} TMG and the
spacelike stretched black hole solution. 
In Sec. \ref{sec:field}, we consider the Klein-Gordon wave equation 
for the massless scalar field under the spacelike stretched black hole
background. Then, general solutions are found to be linear
combinations 
of hypergeometric functions and an appropriate boundary condition is
imposed. The matching coefficients between wave functions near horizon 
and asymptotic regions are identified.
In Sec. \ref{sec:fluxes}, we finally compute the absorption cross
section from the ratio between in-going modes of fluxes near
horizon and that of the asymptotic region. The absorption cross section is composed of two
pieces, the well-known area law and unexpected temperature-dependent
term.  {In Sec. \ref{sec:qn}, a quasinormal mode of a propagating scalar field is obtained by an appropriate quasinormal boundary condition. 
Finally, we summarize and discuss our results in Sec. \ref{sec:discuss}.}

\section{Preliminaries}\label{sec:tmg}

Varying the action (\ref{eq:tmg}) with respect to the metric leads to the bulk equation of motion,
\begin{equation}
\label{eq:eqnmot}
G_{\mu\nu} - \frac{1}{\ell^2}g_{\mu\nu} + \frac{\ell}{3\nu} C_{\mu\nu}=0,
\end{equation}
where the Cotton tensor is
\begin{equation}
C_{\mu\nu} = \epsilon_{\mu}^{~\lambda\sigma}\nabla_{\lambda}\left(R_{\sigma\nu} - \frac{1}{4} g_{\sigma\nu} R\right).
\end{equation}
For the AdS$_3$ geometry, the chiral gravity exists at the
chiral point at $\nu=1/3$ or $\mu\ell=1$
\cite{Li:2008dq}. 
The interesting solution which is free from a naked
CTC is the spacelike stretched solution for $\nu^2 > 1$\cite{Anninos:2008fx, Bouchareb:2007yx, Moussa:2008sj}\footnote{The warped solutions were already discovered in \cite{Israel:2004vv,Detournay:2005fz} as a marginal deformation of the $SL(2,R)$ WZW model in the context of string theory and discussed the connection to the warped solution \cite{Compere:2008cw} and the dual CFT description of TMG with the warped boundary conditions \cite{Compere:2008cv}  by Anninos {\it et. al} \cite{Anninos:2008fx}.}, which is given by
\begin{equation}
(ds)^2 = -N^2(r) dt^2 + \ell^2 R^2(r) (d\theta + N^{\theta}(r)dt)^2 + \frac{\ell^4 dr^2}{4R^2(r)N^2(r)},
\end{equation}
where
\begin{eqnarray}
&& R^2(r) = \frac{r}{4}\left(3(\nu^2-1)r+(\nu^2+3)(r_{+}+r_{-})-4\nu\sqrt{r_{+}r_{-}(\nu^2+3)}\right),\\
&& N^2(r) = \frac{\ell^2(\nu^2+3)(r-r_{+})(r-r_{-})}{4R^2(r)},\\
&& N^{\theta}(r) = \frac{2\nu r-\sqrt{r_{+}r_{-}(\nu^2+3)}}{2R^2(r)}.
\end{eqnarray}

The Killing vectors are defined as $\chi^{a} \equiv \xi^{a} + \Omega_{H}\phi^{a}$ where the angular velocity is 
\begin{equation}
\Omega_{H} \equiv - \left.\frac{g_{t\theta}}{g_{\theta\theta}}\right|_{H} = -\frac{2}{2\nu r_{+}-\sqrt{(\nu^2+3)r_{+}r_{-}}},
\end{equation}
where the subscript $H$ denotes the value at the horizon of $r=r_{+}$.
Then, the Hawking temperature can be found from the surface gravity defined as $\kappa_{H}^2 = - \frac{1}{2}(\nabla_{a}\chi_{b})(\nabla^{a}\chi^{b})|_{H}$, which is
\begin{equation}
T_{H} \equiv \frac{\kappa_{H}}{2\pi} = \frac{(\nu^2+3)(r_{+}-r_{-})}{4\pi (2\nu r_{+}-\sqrt{(\nu^2+3)r_{+}r_{-}})}.
\end{equation}
Note that this black object is closely related to the solution
{originally} discovered in Ref. \cite{Bouchareb:2007yx, Moussa:2008sj}, which is connected with
the coordinate transformation\footnote{This coordinate transformation
  breaks down at the critical point of $\nu = 1$ and the negative
  $\nu$ yields the 
unphysical result. See the appendix in Ref. \cite{Anninos:2008fx} for
more details.} when $\nu^2>1$.

The entropy, ADT mass, and angular momentum are found to be \cite{Bouchareb:2007yx, Abbott:1982jh, Deser:2002rt}
\begin{eqnarray}
&& S=\frac{\pi\ell}{24\nu G_{N}}\left[(9\nu^2+3)r_{+}-(\nu^2+3)r_{-}-4\nu\sqrt{(\nu^2+3)r_{+}r_{-}}\right]\\
&& M_{ADT} = \frac{(\nu^2+3)}{24G_{N}} \left[r_{+}+r_{-}-\frac{\sqrt{(\nu^2+3)r_{+}r_{-}}}{\nu}\right]\\
&& J_{ADT} = \frac{\nu\ell(\nu^2+3)}{96G_{N}}\left[ \frac{24^2 G_{N}^2}{(\nu^2+3)^2}M_{ADT}^2 - \frac{(5\nu^2+3)}{4\nu^2}(r_{+}-r_{-})^2\right],
\end{eqnarray}
respectively, and one can easily confirm the first law of thermodynamics between above
quantities \cite{Anninos:2008fx}. In the next section, we shall
compute the scattering absorption cross section of a scalar field
propagating 
in the spacelike stretched black hole background.

\section{Field Equation and Boundary Condition}\label{sec:field}

We start with the Klein-Gordon equation of the massless scalar field
under the 
spacelike stretched warped AdS$_3$ black hole background,
\begin{equation}
\frac{1}{\sqrt{-g}}\partial_{\mu}\left(\sqrt{-g}g^{\mu\nu}\partial_{\nu}\Phi\right) = 0,
\end{equation}
where $\Phi=\Phi(t,r,\theta)$. 
Using the separation of variables of $\Phi(t,r,\theta)\equiv
\psi(r)e^{-i\omega t+i\mu\theta}$, the radial equation of motion is written as
\begin{eqnarray}
\label{eq:radeqn}
 \psi''(r) + \frac{(2r-r_{+}-r_{-})}{(r-r_{+})(r-r_{-})}\psi'(r) -\frac{\left(\alpha r^2 +\beta r +\gamma \right)}{(r-r_{+})^2(r-r_{-})^2}\psi(r)=0,
\end{eqnarray}
where
\begin{eqnarray}
&& \alpha\equiv -\frac{3\omega^2(\nu^2-1)}{(\nu^2+3)^2},\label{eq:para1}\\
&& \beta \equiv -\frac{\{\omega^2(\nu^2+3)(r_{+}+r_{-})-4\nu(\omega^2\sqrt{r_{+}r_{-}(\nu^2+3)}-2\mu\omega)\}}{(\nu^2+3)^2},\label{eq:para2}\\
&& \gamma \equiv -\frac{4(\mu^2-\omega \mu\sqrt{r_{+}r_{-}(\nu^2+3)})}{(\nu^2+3)^2}\label{eq:para3}
\end{eqnarray}
and the prime denotes a derivative with respect to $r$.

The general solution to Eq. (\ref{eq:radeqn}) 
can be obtained in terms of the second kind 
hypergeometric function $F(a,b,c;z)$,
\begin{eqnarray}
\label{eq:sols}
\psi(r) &=& C_{1}  \left(r-r_{-}\right)^{-\frac{\sqrt{\alpha r_{-}^2+\beta r_{-} +\gamma}}{r_{+}-r_{-}}} \left(r-r_{+}\right)^{\frac{\sqrt{\alpha r_{+}^2+\beta r_{+} +\gamma}}{r_{+}-r_{-}}}F\left[{\mathcal A}_{-},{\mathcal B}_{-},{\mathcal C}_{-};\frac{r_{+}-r}{r_{+}-r_{-}}\right]\nonumber\\&+& C_{2}  \left(r-r_{-}\right)^{-\frac{\sqrt{\alpha r_{-}^2+\beta r_{-} +\gamma}}{r_{+}-r_{-}}} \left(r-r_{+}\right)^{-\frac{\sqrt{\alpha r_{+}^2+\beta r_{+} +\gamma}}{r_{+}-r_{-}}}F\left[{\mathcal A}_{+},{\mathcal B}_{+},{\mathcal C}_{+};\frac{r_{+}-r}{r_{+}-r_{-}}\right],
\end{eqnarray}
where 
\begin{eqnarray}
&&{\mathcal A}_{\pm} = \frac{\mp  \sqrt{\alpha r_{+}^2+\beta r_{+}+\gamma}-\sqrt{\alpha r_{-}^2+\beta r_{-}+\gamma}}{r_{+}-r_{-}} \pm \frac{1}{2}\sqrt{1+4 \alpha}+\frac{1}{2}\label{eq:para4}\\
&&{\mathcal B}_{\pm} = {\mathcal A}_{\pm}\mp\sqrt{1+4 \alpha}\label{eq:para5}\\
&&{\mathcal C}_{\pm} = 1\pm\frac{2\sqrt{\alpha r_{+}^2 + \beta r_{+} +\gamma}}{r_{+}-r_{-}}\label{eq:para6}.
\end{eqnarray}
If we consider the near horizon limit of $r\simeq r_{+}$, the general solution (\ref{eq:sols}) becomes
\begin{eqnarray}
\label{eq:nearsol}
\psi_{near}(r) &\simeq& \hat{C}_{1} \left(r-r_{+}\right)^{\frac{\sqrt{\alpha r_{+}^2+\beta r_{+} +\gamma}}{r_{+}-r_{-}}} + \hat{C}_{2} \left(r-r_{+}\right)^{-\frac{\sqrt{\alpha r_{+}^2+\beta r_{+} +\gamma}}{r_{+}-r_{-}}}\nonumber\\
&=& \hat{C}_{1} {\rm exp}\left[\frac{\sqrt{\alpha r_{+}^2+\beta r_{+} +\gamma}}{r_{+}-r_{-}} \ln (r-r_{+})\right]\nonumber\\
 &&\qquad\qquad\qquad\qquad+\hat{C}_{2} {\rm exp}\left[-\frac{\sqrt{\alpha r_{+}^2+\beta r_{+} +\gamma}}{r_{+}-r_{-}} \ln (r-r_{+})\right],
\end{eqnarray}
where $\hat{C}_{1,2} \equiv
C_{1,2}\left(r_{+}-r_{-}\right)^{-\frac{\sqrt{\alpha r_{-}^2+\beta
      r_{-} +\gamma}}{r_{+}-r_{-}}}$. Note that in the low-frequency
limit of $\omega << 1$, we have the purely imaginary in the exponent since
\begin{equation}
\sqrt{\alpha r_{+}^2+\beta r_{+} +\gamma} = i\left[\frac{2\mu}{(\nu^2+3)} + {\mathcal O}(\omega)\right],
\end{equation}
which gives the ``in-going'' and ``out-going'' coefficients, $C_{in} \equiv {C}_{2}$ and $C_{out} \equiv {C}_{1}$, respectively.

On the other hand, in the asymptotic region of $r\rightarrow \infty$, the wave equation can be written in the form of
\begin{equation}
\psi''(r) + \frac{2}{r} \psi'(r) + \frac{\hat{\alpha} r+\hat{\beta}}{r^3} \psi(r) = 0,
\end{equation}
whose solution is given by the linear combination of the Bessel functions, $J_{\nu}(x)$ and $Y_{\nu}(x)$,
\begin{eqnarray}
\psi_{asym}(r) = \frac{A_{1}}{\sqrt{r}}  {J}_{-\sqrt{1+4\alpha}}\left(2\sqrt{\hat{\beta}r^{-1}}\right)+ \frac{A_{2}}{\sqrt{r}} {Y}_{-\sqrt{1+4\alpha}}\left(2\sqrt{\hat{\beta}r^{-1}}\right)
\end{eqnarray}
where we define $\hat{\alpha}\equiv -\alpha$ and $\hat{\beta}\equiv -\beta$. For fixed $\nu$ and $z\rightarrow 0$, the Bessel functions are expanded in the form of \cite{as}
\begin{equation}
J_{\nu}(z) \sim \frac{1}{\Gamma(\nu+1)} \left(\frac{z}{2}\right)^{\nu},~~Y_{\nu}(z)\sim -\frac{\Gamma(\nu)}{\pi} \left(\frac{z}{2}\right)^{-\nu}.
\end{equation}
Thus the asymptotic solution can be written in the polynomial form of
\begin{equation}
\label{eq:asymsol}
\psi_{asym}(r) \simeq \hat{A}_{1} r^{-\frac{1}{2}+\frac{1}{2}\sqrt{1-4\hat{\alpha}}} + \hat{A}_{2} r^{-\frac{1}{2}-\frac{1}{2}\sqrt{1-4\hat{\alpha}}}
\end{equation}
where the coefficients are 
\begin{equation}
\hat{A}_{1} \equiv \frac{A_{1}}{\Gamma(1-\sqrt{1-4\hat{\alpha}})\hat{\beta}^{\frac{1}{2}\sqrt{1-4\hat{\alpha}}}},~~\hat{A}_{2}\equiv \frac{A_{2}\Gamma(-\sqrt{1-4\hat{\alpha}})}{\pi\hat{\beta}^{-\frac{1}{2}\sqrt{1-4\hat{\alpha}}}}.
\end{equation}

Here, in order to decompose the ``in-going'' and ``out-going'' modes
in the asymptotic solution, 
we define ${A}_{1}\hat{\beta}^{-\frac{1}{2}\sqrt{1-4\hat{\alpha}}}
\equiv A_{in}+A_{out}$ and
${A}_{2}\hat{\beta}^{\frac{1}{2}\sqrt{1-4\hat{\alpha}}} \equiv -i h
(A_{in}-A_{out})$, and then we can rewrite the asymptotic solution in Eq. (\ref{eq:asymsol}) as
\begin{eqnarray}
\label{eq:decomasym}
\psi_{asymp}(r) &\simeq& A_{in}\left(\frac{r^{-\frac{1}{2}+\frac{1}{2}\sqrt{1-4\hat{\alpha}}}}{\Gamma(1-\sqrt{1-4\hat{\alpha}})} -ih \frac{\Gamma(-\sqrt{1-4\hat{\alpha}})}{\pi} r^{-\frac{1}{2}-\frac{1}{2}\sqrt{1-4\hat{\alpha}}}\right)\nonumber\\
&& + A_{out}\left(\frac{r^{-\frac{1}{2}+\frac{1}{2}\sqrt{1-4\hat{\alpha}}}}{\Gamma(1-\sqrt{1-4\hat{\alpha}})}+ih\frac{\Gamma(-\sqrt{1-4\hat{\alpha}})}{\pi}r^{-\frac{1}{2}-\frac{1}{2}\sqrt{1-4\hat{\alpha}}} \right),
\end{eqnarray}
where $h$ is a positive dimensionless  {numerical} constant which will be taken to be independent of the energy $\omega$ \cite{bss,ko}. Note 
that this constant can be chosen so that the absorption cross section can be expressed by the area of the black hole in the low-frequency regime \cite{bss, dgm}. On the other hand, it can be chosen so as to have the usual value of the Hawking temperature \cite{ko} or to make the sum of absorption and reflection coefficients be unity \cite{kko}.
 {This ambiguity comes from the fact that there exists an arbitrary freedom when we decompose the amplitude of the wave function into in-going and out-going modes. However, this freedom can be chosen as a numerical factor by appropriate physical situations.}

Now, we consider the functional transformation of the hypergeometric function \cite{as},
\begin{eqnarray}
F(a,b;c;z) &=& (1-z)^{-a}\frac{\Gamma(c)\Gamma(b-a)}{\Gamma(b)\Gamma(c-a)} F\left(a,c-b;a-b+1;\frac{1}{1-z}\right)\nonumber\\
&&\qquad+(1-z)^{-b}\frac{\Gamma(c)\Gamma(a-b)}{\Gamma(a)\Gamma(c-b)} F\left(b,c-a;b-a+1;\frac{1}{1-z}\right),
\end{eqnarray}
where we define $z\equiv (r_{+}-r)/(r_{+}-r_{-})$, then
we have $z\rightarrow 0$ as $r\rightarrow r_{+}$ while it negatively
diverges when $r\rightarrow\infty$. The above transformation of the
hypergeometric function will be used at the asymptotic region; $1/(1-z)\rightarrow 0$ when $r\rightarrow\infty$.\footnote{In other words, it implies that the behavior of the general solution in terms of the hypergeometric function can be easily controllable at asymptotic region by means of this transformation.} Therefore, Eq. (\ref{eq:sols}) becomes
\begin{eqnarray}
\label{eq:transol}
\psi(r) &=&C_{out}  \left(r-r_{-}\right)^{-\frac{\sqrt{\alpha r_{-}^2+\beta r_{-} +\gamma}}{r_{+}-r_{-}}} \left(r-r_{+}\right)^{\frac{\sqrt{\alpha r_{+}^2+\beta r_{+} +\gamma}}{r_{+}-r_{-}}}\\&&\times\left\{\left(\frac{r-r_{-}}{r_{+}-r_{-}}\right)^{-{\mathcal A}_{-}} \frac{\Gamma({\mathcal C}_{-})\Gamma({\mathcal B}_{-}-{\mathcal A}_{-})}{\Gamma({\mathcal B}_{-})\Gamma({\mathcal C}_{-}-{\mathcal A}_{-})}F\left({\mathcal A}_{-},{\mathcal C}_{-} - {\mathcal B}_{-};{\mathcal A}_{-}-{\mathcal B}_{-}+1;\frac{r_{+}-r_{-}}{r-r_{-}}\right)
\right.\nonumber\\
&&+\left.\left(\frac{r-r_{-}}{r_{+}-r_{-}}\right)^{-{\mathcal B}_{-}} \frac{\Gamma({\mathcal C}_{-})\Gamma({\mathcal A}_{-}-{\mathcal B}_{-})}{\Gamma({\mathcal A}_{-})\Gamma({\mathcal C}_{-}-{\mathcal B}_{-})}F\left({\mathcal B}_{-},{\mathcal C}_{-} - {\mathcal A}_{-};{\mathcal B}_{-}-{\mathcal A}_{-}+1;\frac{r_{+}-r_{-}}{r-r_{-}}\right)
\right\}
\nonumber\\
&+& C_{in}  \left(r-r_{-}\right)^{-\frac{\sqrt{\alpha r_{-}^2+\beta r_{-} +\gamma}}{r_{+}-r_{-}}} \left(r-r_{+}\right)^{-\frac{\sqrt{\alpha r_{+}^2+\beta r_{+} +\gamma}}{r_{+}-r_{-}}}\nonumber\\&&\times\left\{\left(\frac{r-r_{-}}{r_{+}-r_{-}}\right)^{-{\mathcal A}_{+}} \frac{\Gamma({\mathcal C}_{+})\Gamma({\mathcal B}_{+}-{\mathcal A}_{+})}{\Gamma({\mathcal B}_{+})\Gamma({\mathcal C}_{+}-{\mathcal A}_{+})}F\left({\mathcal A}_{+},{\mathcal C}_{+} - {\mathcal B}_{+};{\mathcal A}_{+}-{\mathcal B}_{+}+1;\frac{r_{+}-r_{-}}{r-r_{-}}\right)
\right.\nonumber\\
&&+\left.\left(\frac{r-r_{-}}{r_{+}-r_{-}}\right)^{-{\mathcal B}_{+}} \frac{\Gamma({\mathcal C}_{+})\Gamma({\mathcal A}_{+}-{\mathcal B}_{+})}{\Gamma({\mathcal A}_{+})\Gamma({\mathcal C}_{+}-{\mathcal B}_{+})}F\left({\mathcal B}_{+},{\mathcal C}_{+} - {\mathcal A}_{+};{\mathcal B}_{+}-{\mathcal A}_{+}+1;\frac{r_{+}-r_{-}}{r-r_{-}}\right)
\right\}.\nonumber
\end{eqnarray}

At this stage, we need to impose some boundary conditions under the appropriate physical situations. In general, two-independent boundary conditions can be imposed in this analysis; those are based on the two equivalent pictures of probing scalar fields under the black hole background. The first one is for the classical description of black holes; a black hole can absorb the probing fields but nothing can escape from the event horizon, implying $C_{out}=0$ near the horizon. The second one is for the quantum description of the black hole; the asymptotic observer can detect the quantum radiation emitted from the black hole horizon, implying $A_{in}=0$ at asymptotic region. Since both descriptions are equivalent and independent each other, we shall use the first one in this analysis -- $C_{out} = 0$ for convenience.
Then, for the limit of $r\rightarrow \infty$, Eq. (\ref{eq:transol}) becomes ,
\begin{eqnarray}
\label{eq:transol2}
\psi_{r\rightarrow \infty}(r) &\simeq&C_{in} \left[r^{-\frac{\sqrt{\alpha r_{-}^2+\beta r_{-} +\gamma}}{r_{+}-r_{-}}-\frac{\sqrt{\alpha r_{+}^2+\beta r_{+} +\gamma}}{r_{+}-r_{-}}-{\mathcal A}_{+}}(r_{+}-r_{-})^{{\mathcal A}_{+}} \frac{\Gamma({\mathcal C}_{+})\Gamma({\mathcal B}_{+}-{\mathcal A}_{+})}{\Gamma({\mathcal B}_{+})\Gamma({\mathcal C}_{+}-{\mathcal A}_{+})}\right.\nonumber\\&&\qquad\qquad\left.+r^{-\frac{\sqrt{\alpha r_{-}^2+\beta r_{-} +\gamma}}{r_{+}-r_{-}}-\frac{\sqrt{\alpha r_{+}^2+\beta r_{+} +\gamma}}{r_{+}-r_{-}}-{\mathcal B}_{+}}(r_{+}-r_{-})^{{\mathcal B}_{+}} \frac{\Gamma({\mathcal C}_{+})\Gamma({\mathcal A}_{+}-{\mathcal B}_{+})}{\Gamma({\mathcal A}_{+})\Gamma({\mathcal C}_{+}-{\mathcal B}_{+})}
\right]\nonumber\\
&=& C_{in} \left[(r_{+}-r_{-})^{{\mathcal A}_{+}} \frac{\Gamma({\mathcal C}_{+})\Gamma({\mathcal B}_{+}-{\mathcal A}_{+})}{\Gamma({\mathcal B}_{+})\Gamma({\mathcal C}_{+}-{\mathcal A}_{+})}r^{-\frac{1}{2}-\frac{1}{2}\sqrt{1+4\alpha}}\right.\nonumber\\&&\qquad\qquad\qquad\qquad\left.+(r_{+}-r_{-})^{{\mathcal B}_{+}} \frac{\Gamma({\mathcal C}_{+})\Gamma({\mathcal A}_{+}-{\mathcal B}_{+})}{\Gamma({\mathcal A}_{+})\Gamma({\mathcal C}_{+}-{\mathcal B}_{+})}r^{-\frac{1}{2}+\frac{1}{2}\sqrt{1+4\alpha}}
\right].
\end{eqnarray}
Now, comparing this to Eq. (\ref{eq:decomasym}), it can be easily found that there are matching conditions between the Bogoliuvov coefficients, which is given by
\begin{eqnarray}
&&A_{2}= ih(A_{out}-A_{in})=C_{in}\frac{\pi(r_{+}-r_{-})^{{\mathcal A}_{+}} }{\Gamma(-\sqrt{1+4{\alpha}})}\frac{\Gamma({\mathcal C}_{+})\Gamma({\mathcal B}_{+}-{\mathcal A}_{+})}{\Gamma({\mathcal B}_{+})\Gamma({\mathcal C}_{+}-{\mathcal A}_{+})}\\
&&A_{1}= A_{out}+A_{in}=C_{in} \Gamma(1-\sqrt{1+4{\alpha}})(r_{+}-r_{-})^{{\mathcal B}_{+}} \frac{\Gamma({\mathcal C}_{+})\Gamma({\mathcal A}_{+}-{\mathcal B}_{+})}{\Gamma({\mathcal A}_{+})\Gamma({\mathcal C}_{+}-{\mathcal B}_{+})},
\end{eqnarray}
alternatively
\begin{eqnarray}
A_{in}&=& \frac{C_{in}\Gamma({\mathcal C}_{+})}{2}\left[ \Gamma(1-\sqrt{1+4{\alpha}})(r_{+}-r_{-})^{{\mathcal B}_{+}}\frac{\Gamma({\mathcal A}_{+}-{\mathcal B}_{+})}{\Gamma({\mathcal A}_{+})\Gamma({\mathcal C}_{+}-{\mathcal B}_{+})}\right.\nonumber\\&&\qquad\qquad\qquad\qquad\qquad\left. - i\frac{\pi}{h\Gamma(-\sqrt{1+4{\alpha}})}\frac{(r_{+}-r_{-})^{{\mathcal A}_{+}} \Gamma({\mathcal B}_{+}-{\mathcal A}_{+})}{\Gamma({\mathcal B}_{+})\Gamma({\mathcal C}_{+}-{\mathcal A}_{+})}\right]\label{eq:Ain}\\
A_{out}&=&\frac{C_{in}\Gamma({\mathcal C}_{+})}{2}\left[\Gamma(1-\sqrt{1+4{\alpha}})(r_{+}-r_{-})^{{\mathcal B}_{+}}\frac{\Gamma({\mathcal A}_{+}-{\mathcal B}_{+})}{\Gamma({\mathcal A}_{+})\Gamma({\mathcal C}_{+}-{\mathcal B}_{+})}\right.\nonumber\\&&\qquad\qquad\qquad\qquad\qquad\left.  + i\frac{\pi}{h\Gamma(-\sqrt{1+4{\alpha}})} \frac{(r_{+}-r_{-})^{{\mathcal A}_{+}}\Gamma({\mathcal B}_{+}-{\mathcal A}_{+})}{\Gamma({\mathcal B}_{+})\Gamma({\mathcal C}_{+}-{\mathcal A}_{+})}\right].\label{eq:Aout}
\end{eqnarray}
As is well-known, the frequency mixing
appears through this matching condition. In the next section, we will
calculate the absorption cross section.  

\section{Fluxes and Absorption Cross Section}\label{sec:fluxes}

The absorption (${\mathfrak A}$) and the reflection (${\mathfrak R}$) coefficients are defined by the ratio of ``in-going'' and ``out-going'' fluxes as
\begin{equation}
{\mathfrak A} \equiv \left|\frac{{\mathcal F}_{near}^{in}}{{\mathcal F}_{asym}^{in}}\right|,~~{\mathfrak R} \equiv \left|\frac{{\mathcal F}_{asym}^{out}}{{\mathcal F}_{asym}^{in}}\right|,
\end{equation}
respectively. Note that the definition of the flux is given by
\begin{equation}
{\mathcal F} = \frac{2\pi}{i} (r-r_{+})(r-r_{-})\left[\psi^{*}(r)\frac{\partial}{\partial r} \psi(r) - \psi(r)\frac{\partial}{\partial r} \psi^{*}(r)\right],
\end{equation}
and the
``in-going'' flux in the near horizon is calculated 
by using the near-horizon solution (\ref{eq:nearsol}),
\begin{equation}
{\mathcal F}_{near}^{in} = - \omega|C_{in}|^2\left[4\pi r_{+}\frac{\nu(\nu^2+2)(\nu^2+4)}{(\nu^2+3)} + \frac{32\pi^2\nu}{(\nu^2+3)^2}\left(T_{L}+T_{R}\right)\right].
\end{equation}
The left and right temperatures ($T_{L/R}$) are defined by \cite{Anninos:2008fx}                                                                                                                                                                                                                                                                                                                                                                                                                                                                                                                                                                                                                                                              
\begin{equation}
\label{eq:lrtemp}
T_{R} \equiv \frac{(\nu^2+3)(r_{+}-r_{-})}{8\pi},~~T_{L}\equiv \frac{(\nu^2+3)}{8\pi}\left(r_{+}+r_{-}-\frac{\sqrt{(\nu^2+3)r_{+}r_{-}}}{\nu}\right)
\end{equation}
with the relation associated with the Hawking temperature
\begin{equation}
\frac{1}{T_{H}} = \frac{4\pi\nu}{\nu^2+3}\left(1+\frac{T_{L}}{T_{R}}\right).
\end{equation}

On the other hand, one can find the ``in-going'' flux at the asymptotic region
\begin{equation}
{\mathcal F}_{asym}^{in} = - 4h|A_{in}|^2
\end{equation}
and the absorption coefficient is straightforwardly given as
\begin{equation}
\label{eq:refl}
{\mathfrak A} =\frac{\omega}{4h} \left[4\pi r_{+}\frac{\nu(\nu^2+2)(\nu^2+4)}{(\nu^2+3)} + \frac{32\pi^2\nu}{(\nu^2+3)^2}\left(T_{L}+T_{R}\right)\right] \left|\frac{C_{in}}{A_{in}}\right|^2.
\end{equation}
To compute Eq. (\ref{eq:refl}) explicitly, we consider
parameters defined in Eqs. (\ref{eq:para1})-(\ref{eq:para6})
precisely. Let us define the parameters in
Eqs. (\ref{eq:para4})-(\ref{eq:para6}) as
\begin{eqnarray}
&&{\mathcal A}_{+} \equiv m -i(n_{+}+n_{-})\\
&&{\mathcal B}_{+} \equiv 1-m-i(n_{+}+n_{-})\\
&&{\mathcal C}_{+} \equiv 1-2in_{+},
\end{eqnarray}
where
\begin{equation}
m=\frac{1}{2}+\frac{1}{2}\sqrt{1-4\hat{\alpha}},~~n_{\pm}=\frac{\sqrt{\hat{\alpha}r_{\pm}^2+\hat{\beta}r_{\pm}+\hat{\gamma}}}{r_{+}-r_{-}}
\end{equation}
with $\hat{\alpha}=-\alpha$, $\hat{\beta}=-\beta$, and $\hat{\gamma}=-\gamma$. Then, Eqs. (\ref{eq:Ain}) and (\ref{eq:Aout}) can be written in the form of
\begin{eqnarray}
A_{in}&=& \frac{C_{in}\Gamma(1-2in_{+})}{2}\left[ \Gamma(2m)\frac{(r_{+}-r_{-})^{1-m-i(n_{+}+n_{-})}\Gamma(2m-1)}{\Gamma(m-i(n_{+}+n_{-}))\Gamma(m-i(n_{+}-n_{-}))}\right.\nonumber\\&&\qquad\left. - i\frac{\pi}{h\Gamma(2-2m)}\frac{(r_{+}-r_{-})^{m-i(n_{+}+n_{-})} \Gamma(1-2m)}{\Gamma(1-m-i(n_{+}+n_{-}))\Gamma(1-m+i(n_{+}-n_{-}))}\right]\label{eq:Ains}\\
A_{out}&=& \frac{C_{in}\Gamma(1-2in_{+})}{2}\left[ \Gamma(2m)\frac{(r_{+}-r_{-})^{1-m-i(n_{+}+n_{-})}\Gamma(2m-1)}{\Gamma(m-i(n_{+}+n_{-}))\Gamma(m-i(n_{+}-n_{-}))}\right.\nonumber\\&&\qquad\left. + i\frac{\pi}{h\Gamma(2-2m)}\frac{(r_{+}-r_{-})^{m-i(n_{+}+n_{-})} \Gamma(1-2m)}{\Gamma(1-m-i(n_{+}+n_{-}))\Gamma(1-m+i(n_{+}-n_{-}))}\right].\label{eq:Aouts}
\end{eqnarray}
Note that for $1-4\hat{\alpha}\le 0$, 
$m$ includes an imaginary mode and ${\mathcal A}_{+}-{\mathcal
  B}_{+} = 2m-1 = 2 i {\rm Re}(m)$, which makes $\omega \ge
(\nu^2+3)/2\sqrt{3(\nu^2-1)}$ while 
for $1-4\hat{\alpha}>0$, $m$ is real. Here,
we can easily see that the present calculations are ill-defined at the critical point of $\nu^2=1$.
We expand $m$ with respect to $\omega$, which yields
\begin{equation}
\label{eq:mexp}
m=1 - \frac{\nu^2-1}{(\nu^2+3)^2}\omega^2 - \frac{(\nu^2-1)^2}{(\nu^2+3)^4}\omega^4 - \frac{2(\nu^2-1)^3}{(\nu^2+3)^6}\omega^6 + {\mathcal O}(\omega^8),
\end{equation}
which implies that this analysis breaks down for $\nu^2=1$ since the gamma function $\Gamma(2-2m)$ diverges for $m=1$. If we expand $n_{\pm}$ with respect to $\omega$, then we get
\begin{equation}
\label{eq:nexp}
n_{\pm} = \frac{2\mu}{(\nu^2+3)(r_{+}-r_{-})} + \frac{\nu}{4\pi}\left[\frac{4\pi}{\nu^2+3}\left(1+\frac{T_{L}}{T_{R}}\right)-\frac{1}{T_{R}}\left(r_{+}-(\nu^2+3)^2r_{\pm}\right)\right]\omega + {\mathcal O}(\omega^2)
\end{equation}
with $n_{+}-n_{-}\simeq2\nu(\nu^2+3)$ up to the order of
$\omega$. Note that 
if we consider the s-wave sector ($\mu=0$) of the probing scalar
fields, the first dominant contribution of the imaginary modes can be expressed in terms of the left-and right-handed temperatures.

Keeping Eqs. (\ref{eq:mexp}) and (\ref{eq:nexp}) up to the second leading term, the absorption coefficient of $s$-wave modes is obtained
\begin{equation}
{\mathfrak A}_{\mu=0} \simeq\omega \left[4\pi r_{+}\frac{\nu(\nu^2+2)(\nu^2+4)}{h(\nu^2+3)} + \frac{32\pi^2\nu}{h(\nu^2+3)^2}\left(T_{L}+T_{R}\right)\right].
\end{equation}
Then the absorption cross section $\sigma_{abs}^{\mu=0}$ is
\begin{equation}
\sigma_{abs}^{\mu=0}\equiv \frac{{\mathfrak A}_{\mu=0}}{\omega} = \frac{32\pi^2\nu}{h(\nu^2+3)^2}\left(T_{L}+T_{R}\right)+4\pi r_{+}\frac{\nu(\nu^2+2)(\nu^2+4)}{h(\nu^2+3)} .
\end{equation}
Since the black hole area at horizon is related to the left and right temperatures from Eq. (\ref{eq:lrtemp})
\begin{equation}
{\sf A}_{H} \equiv 2\pi R(r_{+}) = \frac{8\pi^2 \nu}{(\nu^2+3)}(T_{R}+T_{L}),
\end{equation}
we can rewrite the absorption cross section of the s-wave sector as
\begin{equation}
\sigma_{abs}^{\mu=0}\equiv \frac{{\mathfrak A}_{\mu=0}}{\omega} = \frac{{\sf A}_{H}}{h(\nu^2+3)} + \frac{4\pi r_{+}\nu(\nu^2+2)(\nu^2+4)}{h(\nu^2+3)}
\end{equation}
alternatively
\begin{equation}
\sigma_{abs}^{\mu=0}\equiv \frac{{\mathfrak A}_{\mu=0}}{\omega} = {\sf A}_{H} + 4\pi r_{+}\nu(\nu^2+2)(\nu^2+4)
\end{equation}
where $h$ can be chosen so as $h=\frac{1}{(\nu^2+3)}$. Note that the absorption cross section in the cosmological TMG is obtained,
which is proportional to the outer-horizon up to a numerical factor apart from the area of the black hole. 

{At first sight, the absorption cross section looks non-singular
  for $\nu^2=1$, however, 
this is not the case since the present 
analysis breaks down for $\nu^2=1$ as seen in Eq. (\ref{eq:mexp}). There is another way to see this explicitly. We need to rewrite $r_{+}$ in terms of $T_{L}$ and $T_{R}$ of Eq. (\ref{eq:lrtemp}).}
Then the absorption cross section can be rewritten as
\begin{equation}
\label{eq:absT}
\sigma_{abs}^{\mu=0} = {\sf A}_{H} + \frac{64\pi^2\nu^3(\nu^2+2)(\nu^2+4)}{3(\nu^2+3)(\nu^2-1)}\left[1 + \frac{1}{4\nu}\sqrt{(\nu^2+3)\left(1-\frac{3(\nu^2+3)(\nu^2-1)T_{R}}{8\pi\nu^2(T_{L}+T_{R})^2}\right)}\right],
\end{equation}
which shows that the cross section is asymmetric for the left and right temperatures. {Note that the final expression of the result is singular at $\nu^2=1$.}

\section{Quasinormal Modes}\label{sec:qn}

In this section, one can also discuss the quasinormal modes so that 
we can impose the quasinormal boundary condition. The presence of the quasi
normal modes describes the decay of tiny perturbation of black hole at equilibrium \cite{Frolov:1998wf}. 
 {However, in three-dimensional black hole backgrounds, it has been studied that there exists a one-to-one correspondence between quasinormal frequencies and the location of the poles of the retarded correlation function of the perturbations in the dual CFT at boundary \cite{Birmingham:2001pj}. The spectrum of quasinormal modes in the cosmological TMG was studied by the linearized perturbation under the BTZ black hole background \cite{Sachs:2008gt,Lee:2008gt} and for the logarithmic boundary CFT \cite{Sachs:2008yi, Myung:2008dm}.}

The quasinormal boundary condition is given by the solution of the wave function satisfying $\psi_{asym}(r) \rightarrow 0$ at asymptotic region.
Then the quasinormal mode for propagating scalar fields under the spacelike stretched black hole background can be easily read from Eq. (\ref{eq:transol2}) \cite{Birmingham:2001pj,Birmingham:2001hc}, {which is given by the condition of the divergence for the gamma functions in the denominator, yielding {that all arguments of 
$\Gamma({\mathcal B}_{+})$, $\Gamma({\mathcal A}_{+})$, $\Gamma({\mathcal C}_{+}-{\mathcal A}_{+})$, $\Gamma({\mathcal C}_{+} - {\mathcal A}_{+})$ should be $k$, where $k \in  {Z}$,
alternatively
\begin{equation}
i(n_{+}\pm n_{-})\pm m = - k.
\end{equation}}
{Indeed there need not to be the same integer $k$ but the resulting computation shows that it can be reduced to some arbitrary integer $k\in Z$. For example, $k \in Z$ but we also have $k+1 \in Z$.}
Then, we get the left and right quasinormal modes for scalar perturbation in the low-frequency regime,
\begin{equation}
\omega_{L} \simeq - i\frac{(\nu^2+3) k }{2\nu},~~
\omega_{R} \simeq \frac{-4\mu - i{(\nu^2+3)(r_{+}-r_{-})k }}{[\nu(r_{+}+r_{-}) - \sqrt{(\nu^2+3)r_{+}r_{-}}]},
\end{equation}
where $\omega_{L/R}$ are the left-and the right-modes of frequencies in accordance with the boundary CFT description \cite{Birmingham:2001pj}. {Note that this result is valid for any $\nu$}
and the quasinormal frequencies are asymmetric and 
different from that of the AdS$_3$ spacetimes \cite{Sachs:2008gt,Lee:2008gt, Sachs:2008yi, Myung:2008dm}, however, it is difficult to compare our result with the previous results for the BTZ case {directly
because the metric for $\nu=1$ case is different from the BTZ metric.}

\section{Discussions}\label{sec:discuss}
We have studied the scattering amplitudes of a scalar field 
under the warped AdS$_3$ black-hole background in the low-frequency
limit, and computed the corresponding absorption cross section for this massless
scalar field in the s-wave limit. 
As a result, the absorption cross section consists of the expected
area part and the additional deformation. 
Note that this deformation seems to be unexpected 
since the absorption cross section was not given by the area of the
black hole. In fact, the general proof of the universality of area expression 
in the absorption cross section has been done 
for any spherically symmetric geometry in arbitrary dimensions
\cite{Das:1996we}.  
We hope this issue could be clarified in elsewhere.

Finally, as seen before, the {scattering} analysis only holds for $\nu^2>1$\footnote{Of course, the case of $\nu^2<1$ can be analyzed in the same manner. However the $\nu^2<1$ case describes timelike and spacelike squashed vacua, for which we always encounter closed timelike curves at large $r$. So we shall not consider this case here. See Ref. \cite{Anninos:2008fx} for more details.} while it breaks down when $\nu=1$ since the Bogoliuvov coefficients from the matching condition diverges at $\nu=1$, requiring the parallel investigation for $\nu=1$.

\vspace{1cm}
{\bf Acknowledgment}\\
{We would like to thank S. Detournay, D. Grumiller, N. Johansson,  M. I. Park, E. J. Son, and M. Yoon for fruitful comments and many discussions. J. J. Oh would like to thank H. S. Yang for useful comments.
W. Kim was supported by the Korea Science and Engineering Foundation 
(KOSEF) grant funded by the Korea government(MOST) 
(R01-2007-000-20062-0).
J. J. Oh was supported by the Korea Research Council
of Fundamental Science \& Technology (KRCF).}

 
\nc{\PR}[3]{Phys. Rev.  { #1} (#3) #2}
\nc{\NPB}[3]{Nucl. Phys.  { B#1} (#3) #2}
\nc{\PLB}[3]{Phys. Lett.  { B#1} (#3) #2}
\nc{\PRD}[3]{Phys. Rev.  { D#1} (#3) #2}
\nc{\PRL}[3]{Phys. Rev. Lett.  { #1} (#3) #2}
\nc{\PREP}[3]{Phys. Rep.  { #1} (#3) #2}
\nc{\EPJ}[3]{Eur. Phys. J.  { #1} (#3) #2}
\nc{\PTP}[3]{Prog. Theor. Phys.  { #1} (#3) #2}
\nc{\CMP}[3]{Comm. Math. Phys.  { #1} (#3) #2}
\nc{\MPLA}[3]{Mod. Phys. Lett.  { A #1} (#3) #2}
\nc{\CQG}[3]{Class. Quant. Grav.  { #1} (#3) #2}
\nc{\NCB}[3]{Nuovo Cimento  { B#1} (#3) #2}
\nc{\ANNP}[3]{Ann. Phys. (N.Y.)  { #1} (#3) #2}
\nc{\GRG}[3]{Gen. Rel. Grav.  { #1} (#3) #2}
\nc{\MNRAS}[3]{Mon. Not. Roy. Astron. Soc.  { #1} (#3) #2}
\nc{\JHEP}[3]{JHEP  { #1} (#3) #2}
\nc{\JCAP}[3]{JCAP  { #1}, #2 {#3}}
\nc{\ATMP}[3]{Adv. Theor. Math. Phys.  { #1} (#3) #2}
\nc{\AJP}[3]{Am. J. Phys.  { #1} (#3) #2}
\nc{\ibid}[3]{{\it ibid.}  { #1} (#3) #2}
\nc{\ZP}[3]{Z. Physik  { #1} (#3) #2}
\nc{\PRSL}[3]{Proc. Roy. Soc. Lond.  { A#1} (#3) #2}
\nc{\LMP}[3]{Lett. Math. Phys.  { #1} (#3) #2}
\nc{\AM}[3]{Ann. Math.  { #1} (#3) #2}
\nc{\hepth}[1]{{\tt [arxiv:hep-th/{#1}]}}
\nc{\grqc}[1]{{\tt [arxiv:gr-qc/{#1}]}}
\nc{\astro}[1]{{\tt [arxiv:astro-ph/{#1}]}}
\nc{\hepph}[1]{{\tt [arxiv:hep-ph/{#1}]}}
\nc{\phys}[1]{{\tt [arxiv:physics/{#1}]}}

\end{document}